\documentclass[reprint, superscriptaddress, amsmath,amssymb, aps, prresearch, longbibliography, floatfix]{revtex4-2}

\usepackage{amsmath}
\usepackage{graphicx}
\usepackage{ulem}
\usepackage{bm}
\usepackage{natbib}
\usepackage{dsfont}
\usepackage{tikz}
\usepackage{epsfig}
\usepackage{feynmf}
\usepackage{blindtext, rotating}
\usepackage{mathtools}
\usepackage{dsfont}
\usepackage{subcaption}
\usepackage{physics}
\usepackage{amsfonts}
\usepackage{xcolor}
\usepackage{ragged2e}
\usepackage{siunitx}
\usepackage{comment}
\usepackage{soul}
\usepackage{lipsum}
\usepackage{hyperref} 
\hypersetup{breaklinks=true, colorlinks=true, citecolor=blue, linkcolor=cyan, urlcolor=blue,filecolor=blue}

\usepackage{xcolor} 

\DeclareCaptionJustification{justified}{\justifying}

\DeclareSIUnit{\rad}{rad}

\definecolor{bright_blue}{HTML}{85C1E9}
\definecolor{middle_blue}{HTML}{2E86C1}
\definecolor{dark_blue}{HTML}{1B4F72}

\begin{document}

\title{All electrical cooling of an optically levitated nanoparticle}

\author{Oscar Kremer}
\affiliation{Department of Electrical Engineering, Pontifical Catholic University of Rio de Janeiro, 22451-900 Rio de Janeiro, RJ, Brazil}

\author{Igor Califrer}
\affiliation{Department of Physics, Pontifical Catholic University of Rio de Janeiro, Rio de Janeiro 22451-900, Brazil}

\author{Daniel Tandeitnik}
\affiliation{Department of Physics, Pontifical Catholic University of Rio de Janeiro, Rio de Janeiro 22451-900, Brazil}

\author{Jean Pierre von der Weid}
\affiliation{Department of Electrical Engineering, Pontifical Catholic University of Rio de Janeiro, 22451-900 Rio de Janeiro, RJ, Brazil}

\author{Guilherme Tempor\~ao}
\affiliation{Department of Electrical Engineering, Pontifical Catholic University of Rio de Janeiro, 22451-900 Rio de Janeiro, RJ, Brazil}

\author{Thiago Guerreiro}
\email{barbosa@puc-rio.br}
\affiliation{Department of Physics, Pontifical Catholic University of Rio de Janeiro, Rio de Janeiro 22451-900, Brazil}

\begin{abstract}
We implement an all electrical controller for 3D feedback cooling of an optically levitated nanoparticle capable of reaching sub-Kelvin temperatures for the center of mass motion. The controller is based on an optimal policy where state estimation is made by delayed position measurements. The method offers a simplified path for pre-cooling and decoupling the transverse degrees of freedom of the nanoparticle. Numerical simulations show that in an improved setup with quantum limited detection, 3D ground state cooling and all electrical quantum control can be achieved.
\end{abstract}

\maketitle

\section{Introduction}

Optical tweezers \cite{ashkin1986observation} have emerged as a valuable tool for isolating and controlling the motion of micro- and nano-objects \cite{millen2020optomechanics, gonzalez2021levitodynamics, gieseler2021optical}. By clever combinations with electric and magnetic traps and actuators \cite{paul1990electromagnetic, goldwater2019levitated,dania2021optical,bykov2022hybrid,bonvin2023state},  optical traps can be used to design highly sensitive sensors for force, acceleration, and torque \cite{hempston2017force, ranjit2016zeptonewton, ricci2019accurate, ahn2020ultrasensitive, montoya2022scanning}, with a high degree of control enabling cooling of the center-of-mass motion of a levitated nanoparticle to the ground state \cite{delic2020cooling, piotrowski2023simultaneous, magrini2021real, tebbenjohanns2021quantum}.  Moreover, tweezers provide a versatile platform for many-body \cite{rieser2022tunable, penny2023sympathetic, livska2023cold, brandao2021coherent,  vijayan2023cavity}  and fundamental physics experiments, with applications in diverse areas such as stochastic thermodynamics \cite{gieseler2018levitated, dechant2015all, millen2016perspective,debiossac2020thermodynamics, debiossac2022non}, nonlinear dynamics \cite{gieseler2013thermal, ricci2017optically, debiossac2020thermodynamics, kremer2023perturbative, muffato2024generation}, the search for new particles and forces of nature \cite{arvanitaki2013detecting,afek2021limits, monteiro2020search,moore2021searching,afek2022coherent, winstone2022high}, and unprecedented tests of quantum mechanics \cite{pikovski2012probing,weiss2021large,neumeier2024fast, grochowski2023quantum, kamba2023revealing}. All these applications require the levitated object to be well isolated from its surrounding environment, which is mainly limited by the vacuum quality of the experiment, photon recoil heating \cite{jain2016direct}, and black body radiation \cite{hackermuller2004decoherence}. Regarding the vacuum quality, since the nano-object is initially trapped at atmospheric pressure, it is thermalized at room temperature, preventing stable trapping at low pressures and rendering the trapping potential nonlinear due to large thermal fluctuations \cite{gieseler2013thermal}. Therefore, cooling the object's motion is often a prerequisite for levitation experiments.

Active feedback cooling \cite{conangla2019optimal, tebbenjohanns2019cold}, in particular parametric cooling, has emerged as the standard technique for achieving 3D cooling of the levitated nanoparticle's motion \cite{gieseler2012subkelvin}, enabling temperatures as low as sub-mK \cite{jain2016direct}. In practice, parametric control techniques are often used as a precooling mechanism. The performance of parametric feedback, however, comes at the cost of employing a nonlinear control protocol which modulates a portion of the optical trapping power according to the resonance frequencies of the nanoparticle. In addition, expensive electro- (EOM) or acousto-optic (AOM) modulators must be used in combination with
lock-in devices capable of modulating a signal locked to the particle's motion.  Alongside the parametric control, once the thermal occupation number has been reduced to around $ 10^{3} $, the levitated object's charge can be exploited to further control its motion along one direction to even lower temperatures all the way into the quantum ground state \cite{magrini2021real, tebbenjohanns2021quantum}. 

In this letter we explore an all electrical approach  to pre-cool the motion of a levitated nanoparticle from room temperature to a point where the trap's nonlinear features are significantly reduced and stable trapping can be achieved in high-vacuum ($p < $ \SI{e{-3}}{mbar}). To do so, we design a simple electric actuator based on a custom made printed circuit board (PCB), capable of influencing the particle's motion via Coulomb forces. Fine alignment of the PCB with the levitated nanoparticle is not required. After a careful calibration of the electrical forces, we employ a delayed feedback scheme to 3D cool the center-of-mass (CoM) motion of the particle. We experimentally measure the effect of the delay in the feedback force and show excellent agreement with theoretical predictions \cite{rosinberg2015stochastic}. Finally, we successfully demonstrate 3D cooling down to sub-Kelvin temperatures while completely avoiding modulation of the trap's power, in a first step towards the larger effort of simplifying optomechanical cooling experiments. With numerical simulations based on our electrical actuator we argue that, in combination with a stiffer optical trap, quantum-limited detection for all three axes \cite{tebbenjohanns2019optimal} and optimal quantum state estimation \cite{belavkin1995quantum, belavkin1999measurement, wiseman2009quantum}, all electrical 3D ground state cooling can be achieved in our setup.

We highlight that 3D electrical feedback cooling of levitated nanoparticles has been recently implemented in levitated optomechanics experiments -- see \cite{melo2023vacuum, bonvin2023hybrid, gosling2024sensing} for examples using integrated chip photonics, hybrid optical Paul trap and finely aligned electrode tips. Our setup adds a simplified solution to that list, while still offering the possibility of 3D quantum control of a levitated nanoparticle. This paper is organized as follows. In Sec. \ref{sec:theory} we briefly describe the equations of motion and the Linear Quadratic Regulator (LQR), used to evaluate the optimal proportional and derivative gains used in the control feedback. Next, Sec. \ref{sec:experiment} describes the experimental setup, while \ref{sec:results} shows the results on all electrical feedback cooling and the prospects for 3D ground state cooling. We conclude in Sec. \ref{sec:conclusions} with a brief discussion.

\section{Theory}\label{sec:theory}

The CoM motion along the $x$, $y$ and $z$-axes of an optically levitated nanoparticle trapped by a strongly focused Gaussian beam can be effectively modeled through a set of second-order Langevin equations,
\begin{subequations}\label{eq:2.theory-langevin-equation}
    \begin{align}
    \ddot{x}(t) + \gamma_m\dot{x}(t) + \Omega_x^2 x(t) = \frac{1}{m} F_{\text{th},x}(t)+b_xu_x,\\
    \ddot{y}(t) + \gamma_m\dot{y}(t) + \Omega_y^2 y(t) = \frac{1}{m} F_{\text{th},y}(t)+b_yu_y,\\
    \ddot{z}(t) + \gamma_m\dot{z}(t) + \Omega_z^2 z(t) = \frac{1}{m} F_{\text{th},z}(t)+b_zu_z,
    \end{align}
    \label{eq_motion}
\end{subequations}
\noindent where $m$ is the particle's mass, $\gamma_m$ the drag coefficient, $\Omega_i$ the angular frequency along the $i$-axis and $F_{\text{th},i}$ represents the (white-noise) stochastic force on each axis due to residual gas pressure in the vacuum chamber, satisfying
\begin{subequations}\label{eq:2.stochastic_forces}
    \begin{align}
        \langle F_{\text{th},i}(t)\rangle &= 0,\\
        \langle F_{\text{th},i}(t) F_{\text{th}, j}(t+\tau)\rangle &= 2 m\gamma_m k_B T\delta_{ij}\delta(\tau),
    \end{align}
\end{subequations}
\noindent where $k_B$ is the Boltzmann constant, $T$ the residual gas temperature, $\delta_{ij}$ is the Kronecker delta and $i,j \in \{x,y,z\}$. The $b_i u_i$ terms in Eqs. \eqref{eq_motion} account for external forces that may influence the particle's motion, with $u_i$ representing the control signals defining feedback forces acting on the trapped particle. 

By defining the state vector
\begin{equation}
    \mathbf{x}(t) \equiv \begin{bmatrix}
    x(t) &  y(t) & z(t) & \dot{x}(t) &  \dot{y}(t) & \dot{z}(t)
\end{bmatrix}^T,
\end{equation}
one can then write Eqs. \eqref{eq_motion} in the state-variable representation \cite{franklin2015feedback}, resulting in the Multiple-Input-Multiple-Output (MIMO) system
\begin{equation}
    \label{eq:2.state-variable-linear-system}
        \mathbf{\dot{x}}(t) = \mathbf{A}\mathbf{x}(t) + \mathbf{B}\mathbf{u}(t)+ \mathbf{w}(t),
\end{equation}
\noindent where 
\begin{equation}
\mathbf{A} = \begin{bmatrix}
\mathbf{0}_{3\times 3} & \mathbf{I}_3 \\
-\textrm{diag}(\mathbf{\Omega}^2) & -\gamma_m \mathbf{I}_3 \\
\end{bmatrix}
,\,
\mathbf{w}(t) =\frac{1}{m}
 \begin{bmatrix}
\mathbf{0}_{3\times 1} \\
\mathbf{F}_{\rm{th}}(t) \\
\end{bmatrix},
\end{equation}
\noindent and
\begin{equation}
\mathbf{B} = \begin{bmatrix}\mathbf{0}_{3\times 3}\\
            \mathrm{diag}(b_x,b_y,b_z)
            \end{bmatrix}
,\,
\mathbf{u} = \begin{bmatrix} u_x \\ u_y \\ u_z\end{bmatrix},
\end{equation}
with $\mathbf{\Omega}^2=\begin{bmatrix}\Omega_x^2 & \Omega_y^2 & \Omega_z^2 \end{bmatrix}^T$ and $\mathbf{F}_{\rm{th}}(t) = \begin{bmatrix}
    F_{\rm{th}, x}(t)& F_{\rm{th},y}(t) & F_{\rm{th},z}(t)
\end{bmatrix}^T$. Note that due to the geometry of the feedback actuators in our experiment, the submatrix in $ \mathbf{B}$ is not block diagonal, but assumes a more complicated form; see Sec. \ref{sec:experiment} for more details. 

  \begin{figure}[!htb]
      \includegraphics{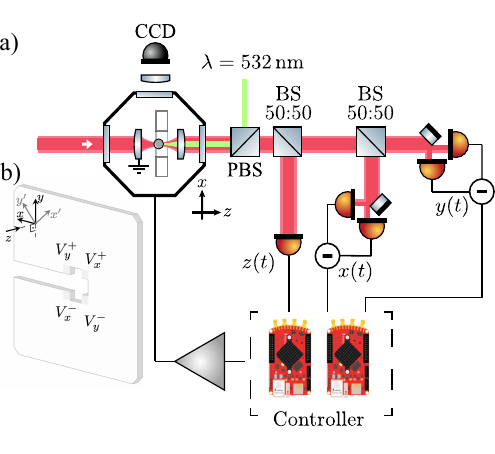}
      \caption{Experimental setup. a) Simplified scheme of the setup. An optical tweezer is assembled within a vacuum chamber, and a CCD is used for imaging of the tweezed particle upon illumination with a \SI{532}{nm} laser beam. The trapping lens is grounded, and detection of forward-scattered light is used to generate the electrical feedback signal sent to the electrodes. The collection lens works as the $z$-electrode, whilst the board shown in b) is placed close to the trap's focus and contains the $x$- and $y$-electrodes. The axes at the top left indicate the orientation between the electrodes' axes ($x^\prime$, $y^\prime$) and the coordinate system of the detection.}
      \label{fig:setup_simplified}
  \end{figure}

Optimal control theory provides tools to find a control policy $\mathbf{u}(t)$ capable of minimizing the energy of a physical system. For linear systems, such as the one described by Eq. \eqref{eq:2.state-variable-linear-system}, this is achieved by the LQR, a controller where the optimization task targets the minimization of a quadratic cost criterion $J$ of the form
\begin{equation}\label{eq:cost-function}
J = \frac{1}{2}\int_0^\infty[\mathbf{x}^T (t)\mathbf{Q}\mathbf{x}(t) + \mathbf{u}^T (t)\mathbf{R}\mathbf{u}(t)]\,dt,
\end{equation}
where $\mathbf{Q}$ is the weighting matrix and $\mathbf{R}$ is the control effort matrix. The optimal control policy which minimizes Eq.\eqref{eq:cost-function} is \cite{kirk2004optimal}
\begin{equation}\label{eq:optimal-control-law}
    \mathbf{u}=-\mathbf{K}\mathbf{x},
\end{equation}
\noindent where $\mathbf{K} = \mathbf{R}^{-1}\mathbf{B}\mathbf{S}$ is the controller's gain matrix and $\mathbf{S}$ is the solution of the algebraic Riccati equation
\begin{equation}
\mathbf{S}\mathbf{A}+\mathbf{A}^T\mathbf{S}+\mathbf{Q}-\mathbf{S}\mathbf{B}\mathbf{R}^{-1}\mathbf{B}^T\mathbf{S}=\mathbf{0}.
\end{equation}

Practical application of the LQR poses the significant challenge of obtaining the complete state vector $\mathbf{x}$. 
Experimentally, access is not granted to $\mathbf{x}$ but rather to a measurement vector $\mathbf{y}$, which is related to the states according to
\begin{equation}
    \mathbf{y}(t) = \mathbf{C}\mathbf{x}(t)+\mathbf{m}(t),
\end{equation}
where $\mathbf{C}$ is known as the output matrix. The term $\mathbf{m}$ is the measurement noise vector and can be expressed as $\mathbf{m}=\begin{bmatrix}\zeta_x(t)&\zeta_y(t)&\zeta_z(t)\end{bmatrix}^T$. Here $\zeta_i(t)$ are zero-mean white-noise processes with variance $\sigma_i^2$, satisfying
\begin{subequations}\label{eq:2.measurement-noise}
    \begin{align}
        \langle \zeta_{i}(t)\rangle &= 0,\\
        \langle \zeta_{i}(t) \zeta_{j}(t+\tau)\rangle &= \sigma_i^2 \delta_{ij}\delta(\tau).
    \end{align}
\end{subequations}

On the one hand, measurements of $x(t)$, $y(t)$ and $z(t)$ can be implemented by collecting forward or backward-scattered light from the nanoparticle \cite{tebbenjohanns2019optimal}. On the other hand, the velocities are not accessible experimentally. An optimal estimation $\mathbf{\hat{x}}$ can be computed by applying real-time filtering techniques to estimate $\mathbf{x}$. For linear dynamics where the disturbances and measurement noises adhere to Eq. \eqref{eq:2.stochastic_forces} and \eqref{eq:2.measurement-noise}, $\mathbf{x}$ is best estimated using the Kalman filter \cite{kalman1960new, aastrom2012introduction}. 

Implementing the Kalman filter significantly increases the complexity of the feedback loop. As a simplification, it is possible to estimate the velocity as being proportional to a delayed position measurement. This approach has proven successful for cooling one of the spatial degrees of freedom of the levitated nanoparticle \cite{conangla2019optimal}, albeit increasing the minimal effective temperature achievable. The effective temperature for each axis can be computed by using the integral \cite{tebbenjohanns2021quantum}
\begin{equation}\label{eq:t_eff}
    T_{\rm{eff}}^i = \frac{m\Omega_i^2}{k_B}\int_0^\infty \Bigg(1+\frac{\Omega^2}{\Omega_i^2}\Bigg)S_{ii}(\Omega)\, d\Omega - \frac{1}{2},
\end{equation}
where $S_{ii}$ is the double-sided Power Spectral Density (PSD) for the particle's motion along the $i$-axis, expressed as
\begin{equation}\label{eq:psd}
    S_{ii}=\frac{2\gamma_m k_B T}{m[(\Omega^2-\Omega_i^2)^2+\gamma_m^2\Omega_i^2]}.
\end{equation}

\section{Experiment}\label{sec:experiment}

The experimental setup is schematically illustrated in Fig. \ref{fig:setup_simplified}a. A CW laser at \SI{1550}{nm} (RIO Orion) amplified by an Erbium-doped fiber amplifier (Keopsys CEFA-C-BO-HP-SM) is used to produce a high-quality Gaussian beam linearly polarized along the $x$ direction with a power of $P_t\approx \SI{2}{W}$, at the output of a single-mode fiber. The beam is focused by an aspheric lens (Thorlabs C330TM-C, NA = 0.68) assembled inside a vacuum chamber, allowing for stable optical trapping. The light scattered by the particle along the forward direction is collimated by a collecting lens (Thorlabs C110TM-C, NA = 0.40). Silica nanoparticles (diameter \SI{143}{nm}, MicroParticles GmbH) are loaded into the vacuum chamber by a nebulizer and trapped at atmospheric pressure. The trapped particle oscillates with resonance frequencies along the three axes given by $\Omega_x/2\pi=\SI[parse-numbers=false]{96.24}{k\hertz}$, $\Omega_y/2\pi=\SI[parse-numbers=false]{101.49}{k\hertz}$ and $\Omega_z/2\pi=\SI[parse-numbers=false]{31.52}{k\hertz}$.

\begin{figure}[t!]
    \centering
    \includegraphics[width=\linewidth]{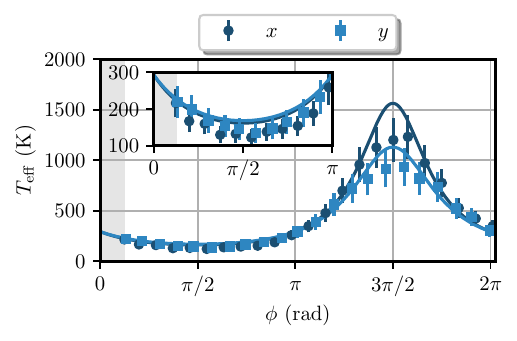}
    \caption{Effect of delayed feedback forces. Comparison between experimental results and theory (solid lines) is presented. Measurements were conducted at room temperature (\SI{293}{K}) and a pressure of \SI{1.2}{mbar}. Each data point corresponds to 10,000 $\SI{50}{ms}$-traces. The used gains were $G_x=\SI[parse-numbers=false]{(9.17\pm0.98)\times 10^{-9}}{\newton/\meter}$ and $G_y=\SI[parse-numbers=false]{(8.97\pm0.97)\times 10^{-9}}{\newton/\meter}$. The gray shaded area marks the region that could not be measured due to the minimal delay imposed by the electronics. The horizontal axis, $\phi$, represents the phase $\Omega_i \tau_i$ introduced by the delay. In the inset, the interval where the delay induces cooling is presented with more detail.} 
    \label{fig:delay-characterization}
\end{figure}

Detection of transversal motion, $x(t)$ and $y(t)$, is carried out using balanced photodiodes (Newport 2117-FC), while information about the longitudinal $z(t)$ direction is obtained by direct intensity photodetection. 
The optical trap is characterized through measurements of the particle's position PSDs for each direction. Information on the occupation numbers and effective temperatures of each direction can also be obtained from the PSDs by using Eq. \eqref{eq:t_eff}.

A PCB containing two orthogonal pairs of electrodes, illustrated in Fig. \ref{fig:setup_simplified}b), is placed in the vicinity of the optical trap's focus, allowing for two-dimensional electrical feedback control of the nanoparticle's CoM motion.
The PCB is designed to be compatible with cage plate optical systems (Thorlabs SP02). Note also that only coarse alignment of the PCB with respect to the levitated nanoparticle is required, and this can be achieved by placing the PCB near the optical focus. Due to the employed control method, coupling between degrees of freedom in the transverse plane is compensated by the calibration process.

\begin{figure*}
    \centering
    \includegraphics{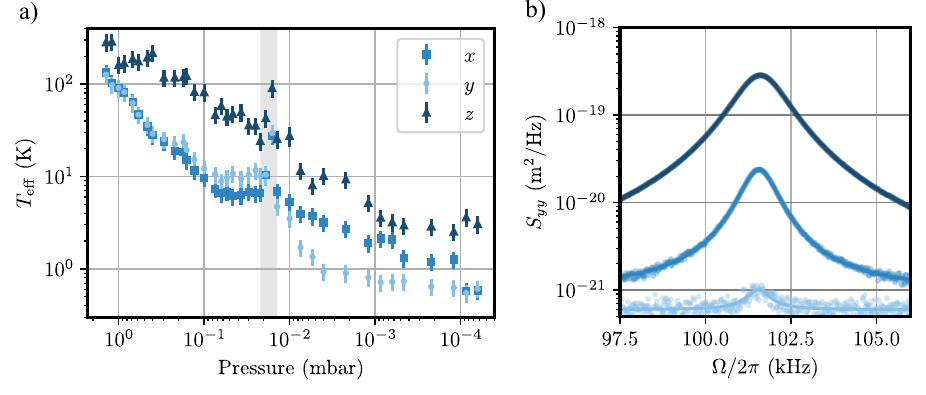}
    \caption{All electrical cooling. a) Dependence between pressure and $x$, $y$ and $z$ effective temperatures. The grey shaded region shows a region of instability, as discussed in the main text. b) PSD of the y motion. Measurements were made at $\SI{1.0}{mbar}$(\protect\tikz[baseline]{\protect\draw[dark_blue, line width=0.5mm] (0,.8ex)--++(0.5,0) ;}), $\SI[parse-numbers=false]{5.4\times 10^{-2}}{mbar}$ (\protect\tikz[baseline]{\protect\draw[middle_blue, line width=0.5mm] (0,.8ex)--++(0.5,0);}) and $\SI[parse-numbers=false]{1.2\times 10^{-4}}{mbar}$ (\protect\tikz[baseline]{\protect\draw[bright_blue, line width=0.5mm,] (0,.8ex)--++(0.5,0) ;}).}
    \label{fig:Cooling-results}
\end{figure*}

A third pair of electrodes is implemented by applying an electric signal to the mount of the collection lens, producing a voltage difference with respect to the grounded trapping lens. Simulations conducted using the finite-elements method have numerically demonstrated that this voltage difference establishes a uniform electric field near the particle's position \cite{kremer2023perturbative}. The signal from the detection is digitally processed by two FPGAs (STEMlab 125-14, Red Pitaya) and analogically amplified before being fed back to the electrodes. We remove any cross-talk between $z$ and $xy$ electrodes by digital filtering, which is facilitated by the difference in characteristic frequencies between the longitudinal and transversal degrees of freedom. Taking this and the geometry of the actuators into consideration, the gain matrix assumes a block diagonal form,
\begin{equation}
    \mathbf{K}=\begin{bmatrix}
        \mathbf{K}_{p,xy} & 0 & \mathbf{K}_{d,xy} & 0\\
        \mathbf{0}_{1\times 2} & k_{p,z} &  \mathbf{0}_{1\times 2} &  k_{d,z}          
    \end{bmatrix}.
\end{equation}

A detailed description of the analogical amplification and the digital processing of the detection signal can be found in Appendix \ref{electronicsAppendix}. Digital processing includes frequency filtering, delaying and application of derivative/proportional gains to the signal. The choice of optimal gains was based on control theory, as presented in Appendix \ref{optimalcontrolAppendix}. Since the theory predicts only a weak dependence of optimal gain on pressure, we consider a single gain to be optimal throughout the experiment.

Appropriate calibration of the electrodes accounts for misalignment between the electrodes' axes and the mechanical modes, allowing for a partial reconstruction of the $\mathbf{B}$ matrix, which assumes a $ 45^{\circ}$ rotated form with respect to the diagonal matrix given by Eqs. \eqref{eq_motion}. During calibration, the effect of the $z$-electrode was observed to be too weak, such that only the $x$- and $y$-electrodes could be calibrated. This has led to applying the control LQR only to the $x$ and $y$ motion and a cold damping protocol \cite{tebbenjohanns2019cold, tebbenjohanns2021quantum} along the $z$ direction ($k_{p,z}=0$). We refer to Appendix \ref{calibrationAppendix} for more information on the calibration procedure.

\section{Results}\label{sec:results}

Proper implementation of the control method as previously described requires precisely delaying each detection signal. The delay characterization process involves applying a force proportional to the delayed position independently in the $x$ and $y$ directions. For instance, referring to Eq. \eqref{eq_motion}, this translates to $u_x=G_{x} x(t-\tau_x)$ for the $x$ coordinate (and similarly for $y$ and $z$). Each delay $\tau_i$ consists of two components, the intrinsic electronic delay $\tau_{e,i}$, and an adjustable delay $\tau_{c,i}$. Fig. \ref{fig:delay-characterization} shows measures of $T_{\rm{eff}}^x$ and $T_{\rm{eff}}^y$ while subjecting the particle to the delayed force. The controllable delay $\tau_{c,i}$ was varied to span the range of $\tau_i$ from $\tau_{e,i}$ to one period of oscillation ($\phi=2\pi$). The experimental results show excellent agreement with the theoretical predictions from \cite{rosinberg2015stochastic}. Furthermore, this measurement allowed for the characterization of the electronic delays, $\tau_{e,x}$ and $\tau_{e,y}$, both of which were determined to be $\SI[parse-numbers=false]{0.639}{\mu\second}$. We assume $\tau_{e,z}$ has the same value.

Figure \ref{fig:Cooling-results}.a) shows the results of 3D feedback cooling. The minimal effective temperatures achieved in the experiment are $T_{\rm{eff}}^x=\SI[parse-numbers=false]{(0.58\pm0.12)}{K}$, $T_{\rm{eff}}^y=\SI[parse-numbers=false]{(0.55\pm0.11)}{K}$ and $T_{\rm{eff}}^z=\SI[parse-numbers=false]{(3.63\pm0.77)}{K}$, for each of the three axes. The gray shaded area in Fig. \ref{fig:Cooling-results}a) depicts an instability region observed near $\SI[parse-numbers=false]{10^{-2}}{mbar}$, characterized by a sudden increase in $T_{\rm{eff}}^i$. We attribute this phenomenon to variations on the net charge of the nanoparticle \cite{ricci2022chemical}. The net charge acts as a linear parameter affecting the input matrix, thus linearly impacting the control gain. As electrode calibration was performed at high pressure ($>\SI{1}{mbar}$), for pressures smaller than $\SI{0.01}{mbar}$, it cannot be assumed that the applied gain was optimal. Nonetheless, stable cooling has been implemented by using only electrical actuators and the application of LQR returned a gain matrix capable of handling any coupling between degrees of freedom in the dynamics. The PSD of the CoM motion for the $y$ direction under three distinct pressures is shown in Fig. \ref{fig:Cooling-results}.b). Feedback cooling not only reduces the area of the PSD, from which the effective temperatures are estimated, but also introduces a term which increases its linewidth, as expected due to the presence of derivative terms in the nanoparticle's motion.

For pressures smaller than $\SI{0.01}{mbar}$, no instability has been encountered, agreeing with results previously shown \cite{frimmer2017controlling}. Therefore, the control protocol employed should be capable of successfully controlling the nanoparticle until the stochastic thermal force becomes negligible and the dynamics starts to be dominated by measurement back-action and photon recoil heating. When compared to parametric cooling, an all electrical approach is advantageous since it avoids contamination of the signal by spurious modulation signals, which are rendered unnecessary. Additionally, in contrast to parametric cooling, the LQR employs a linear control law, thus not affecting the overall linearity of the system. 

\begin{figure}[!htb]
    \centering
    \includegraphics{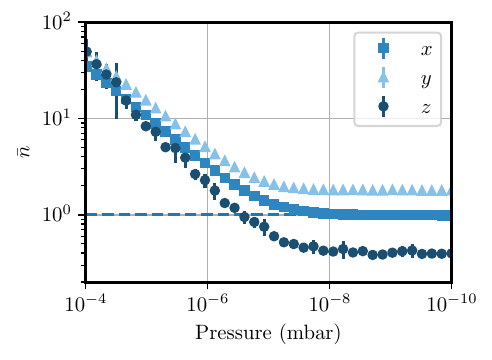}
    \caption{Simulation of optimal all electrical 3D cooling with improved trapping lens and detection scheme: expected thermal occupation numbers, $\bar{n}$, as a function of pressure for the $ x, y$ and $ z $ directions. Dashed line marks a single phonon. Error bars correspond to one standard deviation over 30 simulation runs.}
    \label{fig:3d-ground-state}
\end{figure}

Since the LQR has been successfully employed in combination with Kalman filter for ground-state cooling along the longitudinal axis \cite{magrini2021real}, extending its application as a 3D quantum control policy should be experimentally achievable. By considering the electrode parameters presented in Appendix \ref{calibrationAppendix} and the trapping and detection efficiency parameters reported in \cite{magrini2021real}, we numerically simulated 3D all electrical cooling of a trapped nanoparticle. Figure \ref{fig:3d-ground-state} presents the expected final mean occupation numbers with our all electrical controller. To account for quantum effects, the same parameters of measurement uncertainty, detection efficiency $\eta_z$ and backaction provided in \cite{magrini2021real} were taken into account in the simulation. Note that the simulation considers a backward detection scheme, resulting in a higher detection efficiency for the longitudinal axis compared to the transversal axes, thereby leading to a smaller thermal occupancy for $z$. In contrast, the experiment employed a forward detection scheme, therefore yielding the opposite effect due to limited detection efficiency \cite{tebbenjohanns2019optimal}. The detection efficiency along the transverse axes, $\eta_x$ and $\eta_y$, were computed by considering the expected proportion between the efficiency along $x$ and $y$ and the longitudinal directions for the corresponding trap's NA \cite{tebbenjohanns2019optimal}. For pressures on the order of \SI{1e{-10}}{mbar}, the simulation results agree with the experimental findings in \cite{magrini2021real}. It must be noted that, for higher pressures, we expect that experimental imperfections increase the minimum number of phonons . Moreover, while it is evident that in simulation the thermal occupancy for $y$ exceeds that of $x$, the experimental results in Fig. \ref{fig:Cooling-results}.a) shows the opposite. This most likely arises from experimental imperfections due to detection efficiency in the $x$-axis.

\section{Conclusions}\label{sec:conclusions}

In conclusion, we have demonstrated an all electrical feedback cooling scheme for reducing the CoM temperature of a levitated nanoparticle in high vacuum. Through a simple custom-designed electrical actuator, we have shown sub-Kelvin temperatures for the transverse directions of motion, avoiding the use of nonlinear feedback cooling schemes such as parametric feedback cooling. This greatly simplifies levitated optomechanics experiments by avoiding the need for modulation of the trapping power. Numerical simulations point that future improvements over our setup, in particular implementation of a higher NA trapping lens and of the optimal backward detection scheme reported in \cite{magrini2021real}, should enable all electrical 3D cooling near the ground state, reaching thermal occupation numbers below unity.

\section*{Acknowledgments}

We acknowledge Bruno Melo and Felipe Almeida for useful discussions. T.G. acknowledges the Coordena\c{c}\~ao de Aperfei\c{c}oamento de Pessoal de N\'ivel Superior - Brasil (CAPES) - Finance Code 001, Conselho Nacional de Desenvolvimento Cient\'ifico e Tecnol\'ogico (CNPq), Funda\c{c}\~ao de Amparo \`a Pesquisa do Estado do Rio de Janeiro (FAPERJ Scholarship No. E-26/200.252/2023)  and Funda\c{c}\~ao de Amparo \`a Pesquisa do Estado de São Paulo (FAPESP process No. 2021/06736-5). This work was supported by the Serrapilheira Institute
(grant No. Serra – 2211-42299). 

\bibliography{main}

\newpage

\appendix

\section{Digital control theory}\label{optimalcontrolAppendix}

In the study and analysis of physical systems, time is conventionally treated as a continuous variable. However, when employing signal processing and control methods, a transition to a discrete representation becomes necessary. This is specially crucial when implementing systems on microprocessors or FPGAs, where a set of instructions is executed based on a sampling frequency $f_s$ \cite{franklin1998digital}. The discrete-time formulation of a state-space model can be obtained through the integral approximation, which is based on the assumption that $\mathbf{x}$ and $\mathbf{u}$ remain constant during a sampling period $T_s=1/f_s$. The system evolution is then considered to unfold at fixed time-steps $t_n=nT_s$, leading to the following recursive equations
\begin{subequations}\label{eq:3.2.1.complete-discrete-dynamcis}
    \begin{align} 
        \mathbf{x}_{n+1} &= \mathbf{A}_d \mathbf{x}_n + \mathbf{B}_d\mathbf{u}_n + \bar{\mathbf{w}}_{n}, \\
        \mathbf{y}_{n+1} &= \mathbf{C}_d \mathbf{x}_{n+1} + \bar{\mathbf{m}}_{n+1},
    \end{align}
\end{subequations}
where $\mathbf{A}_d$, $\mathbf{B}_d$ and $\mathbf{C}_d$ can be expressed in terms of their continuous analogues,
\begin{subequations}
\begin{align}
    \mathbf{A}_d &= \sum_{k=0}^\infty \frac{T_s^k}{k!}\mathbf{A}^k,  \\
    \mathbf{B}_d &= (\mathbf{A}_d-\mathbf{I})\mathbf{A}^{-1}\mathbf{B},\\
    \mathbf{C}_d &= \mathbf{C}.
\end{align}
\end{subequations}
Also, $\mathbf{x}_{n}=\mathbf{x}(nT_s)$ and $\mathbf{u}_n=\mathbf{u}(nT_s)$. The discrete disturbance and noise terms, $\bar{\mathbf{w}}_{n}$ and $\bar{\mathbf{m}}_{n}$, represent discrete-time white-noise processes adhering to conditions akin to those established in Eqs. \eqref{eq:2.stochastic_forces} and \eqref{eq:2.measurement-noise} in the main text. Considering $
\bar{\mathbf{w}}_{n} =\frac{1}{m}
 \begin{bmatrix}
\mathbf{0}_{3\times 1}&
\bar{\mathbf{F}}_{\rm{th}, n}
\end{bmatrix}^T
$, with  $\bar{\mathbf{F}}_{\rm{th},n} = \begin{bmatrix}
    \bar{F}_{\rm{th}, x, n}& \bar{F}_{\rm{th},y, n} & \bar{F}_{\rm{th},z,n}
\end{bmatrix}^T$ and $\bar{\mathbf{m}}_{n}=\begin{bmatrix}
    \bar{\zeta}_{x,n}&\bar{\zeta}_{y,n}&\bar{\zeta}_{z,n}
\end{bmatrix}^T$, the conditions are
\begin{subequations}
    \begin{align}
        \langle \bar{F}_{\rm{th},i,k}\rangle &= 0,\\
        \langle \bar{F}_{\rm{th},i,k} F_{\rm{th}, j,k^\prime}\rangle &= 2 m\gamma_m k_B T T_s\delta_{ij}\delta_{kk^\prime},
    \end{align}
\end{subequations}
and 
\begin{subequations}
    \begin{align}
        \langle \bar{\zeta}_{i,k}(t)\rangle &= 0,\\
        \langle \bar{\zeta}_{i,k}\bar{\zeta}_{j,k^\prime}\rangle &= \frac{\sigma_i}{T_s}\delta_{ij}\delta_{kk^\prime}.
    \end{align}
\end{subequations}

Similar to its continuous version, the LQR for discrete-time systems returns an optimal control law, expressed as a linear combination on the states $\mathbf{x}_n$,
\begin{equation}\label{eq:discrete-control-law}
    \mathbf{u}_n = -\mathbf{K}_d\mathbf{x}_n,
\end{equation}
however, the expression for the controller's gain changes to
\begin{equation}\label{eq:discrete-control-gain}
\mathbf{K}_d = (\mathbf{R}_d+
    \mathbf{B}_d^T\mathbf{S}_d\mathbf{B}_d)^{-1}\mathbf{B}_d^T\mathbf{S}_d\mathbf{A}_d,
\end{equation}
where $\mathbf{S}_d$ is the solution of the discrete algebraic Ricatti equation
\begin{equation}
\begin{split}
\mathbf{S}_d&=\mathbf{A}_d^T\mathbf{S}_d\mathbf{A}_d+\mathbf{Q}_d \\
    &-\mathbf{A}_d^T\mathbf{S}_d\mathbf{B}_d(\mathbf{R}_d+\mathbf{B}_d^T\mathbf{S}_d\mathbf{B}_d)^{-1}\mathbf{B}_d^T\mathbf{S}_d\mathbf{A}_d,  
\end{split}
\label{eq:dare}
\end{equation}
and $\mathbf{Q}_d$ and $\mathbf{R}_d$ are the matrices defining the cost function $J_d$ for the digital control law, which reads
\begin{equation}
    J_d = \frac{1}{2}\sum_{n=0}^\infty [\mathbf{x}_n^T\mathbf{Q}_d\mathbf{x}_n + \mathbf{u}_n^T\mathbf{R}_d\mathbf{u}_n].
\end{equation}

\section{Electronic setup}\label{electronicsAppendix}

The control law defined in Eq. \eqref{eq:discrete-control-law} was implemented using two Red Pitayas, each equipped with a Xilinx Zynq 7010 FPGA and a 2 channel 14-bits ADC, allowing for a maximum sampling frequency of $\SI[parse-numbers = false]{125}{M\hertz}$ for two distinct inputs, $x_a$ and $x_b$. The feedback loop incorporated a decimation block, increasing the sampling time $T_s$ from $\SI[parse-numbers=false]{8.00}{n\second}$ to $\SI[parse-numbers=false]{64.00}{n\second}$, enabling synchronous execution of more complex tasks.

\begin{figure*}[!htb]
    \centering
    \includegraphics[width=\textwidth]{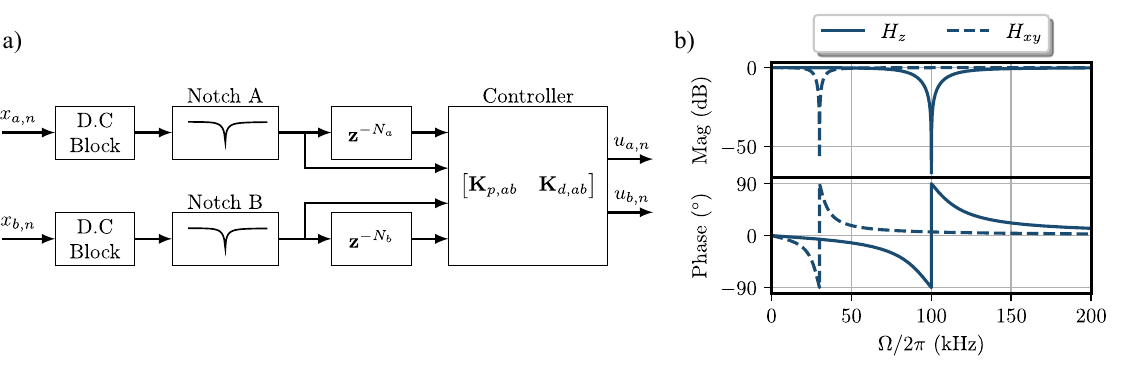}
    \caption{Digital electronic implementation. a) Block diagram illustrating the FPGA implementation for stable control of the particle CoM motion. The digital filters are responsible for signal conditioning. A Block Random Access Memory allows the implementation of delay blocks, delaying the signal in multiples ($N_a$, $N_b$) of the sampling time. The delayed and non-delayed filtered signals are then transmitted to the controllers to compute the output signals. b) Bode plots for each notch filter $  H_{z}$ and $ H_{xy} $, depicting their magnitude and phase behavior for the frequency range of interest.}
    \label{fig:fpga-diagram}
\end{figure*}

In Fig. \ref{fig:fpga-diagram} a simplified block diagram of the main components implemented within each FPGA is shown. The controller block is responsible for computing the output signal $u_{a,n}$ and $u_{b,n}$, being equivalent to the following expression
\begin{equation}\label{eq:generic-control-law-implemented}
    \begin{bmatrix}
        u_{a,n}\\
        u_{b,n}
    \end{bmatrix}=
    \begin{bmatrix}
        k_{p,aa}^d&k_{p,ab}^d&k_{d,aa}^d&k_{d,ab}^d   \\
        k_{p,ba}^d&k_{p,bb}^d&k_{d,ba}^d&k_{d,bb}^d 
    \end{bmatrix}
    \begin{bmatrix}
        \tilde{x}_{a,n}\\
        \tilde{x}_{b,n}\\
        \tilde{x}_{a,n-N_a}\\
        \tilde{x}_{b,n-N_b}
    \end{bmatrix}.
\end{equation}
The signals $\tilde{x}_{a,n}$, $\tilde{x}_{b,n}$ result from passing the inputs through a D.C block and a notch filter, both implemented by using digital biquadratic filters. The constant $k_{p,ij}^d$ and $k_{d,ij}^d$ refer to the digital proportional and derivative gains. The signals $\tilde{x}_{a,n-N_a}$, $\tilde{x}_{b,n-N_b}$ are the delayed positions, serving as estimates of the particle's velocity. 

The notch filter transfer function is shown in Fig. \ref{fig:fpga-diagram}.b). For the FPGA processing the $x$ and $y$ signals, the transfer function used was $H_{xy}$ to remove harmonic components near $\Omega_z$. In the other FPGA, a filter $H_z$ was applied to remove any components sufficiently close to $\Omega_x$ and to $\Omega_y$. The filter's impact on the phase of each signal is approximately constant near each resonance frequency, being included in the overall intrinsic delay of the electronic setup, already described in Section \ref{sec:results}. The computed control signals were sent to non-inverting analog amplifiers, providing a constant gain $A=\SI[parse-numbers = false]{5.00}{\volt / \volt}$ with minimal phase impact for signals with harmonic components from D.C up to $\SI[parse-numbers=false]{150.00}{k\hertz}$.

\section{Model parameters}\label{calibrationAppendix}

Implementation of LQR relies on the accurate extraction of the $\mathbf{A}$ and $\mathbf{B}$ matrices, essential for the correct computation of $\mathbf{A}_d$ and $\mathbf{B}_d$. This appendix clarifies how the parameters that allow the reconstruction of these matrices were extracted for the experiment.

\subsection{Detector calibration}

Assuming the trapped nanoparticle reaches thermal equilibrium with the residual gas in the vacuum chamber, its initial effective temperature along the three axes is approximately $\SI[parse-numbers=false]{293}{\kelvin}$. Calibration of the detection system involves establishing the linear relationship between the PSD of the detector output for motion along the $i$-axis, denoted as $S_{V_iV_i}(\Omega)$, and the displacement PSD for the same axis, denoted as $S_{ii}(\Omega)$ \cite{hebestreit2018calibration},
\begin{equation}\label{eq:detec_calibration}
S_{V_iV_i}(\Omega)= (C_{Vm}^i)^2S_{ii}(\Omega),
\end{equation}
with $C_{Vm}^i$ representing the calibration factor and $S_{V_iV_i}$ being defined by the Lorentzian function,
\begin{equation}\label{eq:voltage-psd}
S_{V_iV_i}(\Omega)= (C_{Vm}^i)^2\frac{2\gamma_m k_B T}{m[(\Omega^2-\Omega_i^2)^2+\gamma_m^2\Omega_i^2]}
\end{equation}

Calibration was done by collecting 10,000 traces, each with a duration of $\SI[parse-numbers=false]{50}{m\second}$. The average PSDs were then fitted to Equation \eqref{eq:voltage-psd}, enabling the extraction of $C_{Vm}^i$, $\Omega_i$ and $\gamma_m$. The coefficients were found to be $$C_{Vm}^{x}= \SI[parse-numbers = false]{(6.87\pm 0.72)\times 10^{5}}{\volt / \meter}$$ $$C_{Vm}^{y}= \SI[parse-numbers = false]{(7.08\pm 0.75)\times 10^{5}}{\volt / \meter}$$ $$C_{Vm}^{z}= \SI[parse-numbers = false]{(1.07\pm 0.11)\times 10^{6}}{\volt / \meter}$$ 

\subsection{Electrodes calibration}

To compute the controller's gain matrix $\mathbf{K}_d$ described in Appendix \ref{optimalcontrolAppendix}, it is necessary to measure the transduction coefficient $C_{NV}^{ij}$ that provides the linear relation between the applied voltage across the electrodes $j$ and the resulting force along the $i$-axis. From these, it is possible to reconstruct the terms of the $\mathbf{B}$ matrix, which due to the geometry of the actuators couples the $x$ and $y$-axes. 

\begin{figure*}[!htb]
    \centering
    \includegraphics{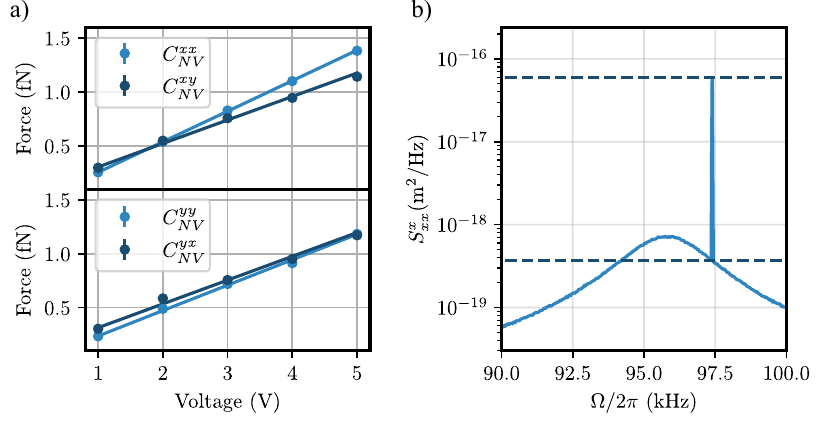}
    \caption{Electrode calibration. a) Calibration curves are presented for each coefficient of the $xy$ plane. Each point corresponds to the analysis of 7000 traces with an individual duration of $\SI{50}{m\second}$. The particle was driven with a sinusoidal signal at $\Omega_{dr}/2\pi=\SI[parse-numbers=false]{97.50}{k\hertz}$. b) PSD of the particle's CoM motion under the action of a sinusoidal force. The dashed line delineates the peak region from which the amplitude of the force $F_0$ can be extracted.}
    \label{fig:calibration_panel}
\end{figure*}

Force calibration of the electrodes can be carried out by measuring the particle's response to sinusoidal voltage drives applied to an individual pair of electrodes at known frequencies near each resonance \cite{ricci2019accurate}. The driving voltage in the electrode $j$ introduces a sinusoidal force $F_{i}^{j}\cos(\Omega_{dr}t)$ which can be observed within the PSD of the driven CoM motion of the $i$ direction $S_{ii}^j$,
\begin{equation}
    S_{ii}^j =  S_{ii}(\Omega) + S_{ii}^{j, el}(\Omega),
\end{equation}
where $S_{ii}(\Omega)$ follows Eq. \eqref{eq:psd} and $S_{ii}^{j,el}(\Omega)$ is
\begin{equation}
        S_{ii}^{j, el}(\Omega)= \frac{F_i^{j2}\tau_{el}\,\textrm{sinc}^2[(\Omega-\Omega_{dr})\tau_{el}]}{m^2[(\Omega^2-\Omega_i^2)^2+\gamma_m^2\Omega^2]},
\end{equation}
\noindent with $\tau_{el}$ being the duration of the measure. 

In Figure \ref{fig:calibration_panel}a), the calibration curves for each coefficient is shown, yielding $$C_{NV}^{xx}= \SI[parse-numbers = false]{(2.83\pm 0.14)\times 10^{-16}}{\newton / \volt}$$ $$C_{NV}^{xy}= \SI[parse-numbers = false]{(2.18\pm 0.13)\times 10^{-16}}{\newton / \volt}$$ $$C_{NV}^{yx}= \SI[parse-numbers = false]{(2.21\pm 0.13)\times 10^{-16}}{\newton / \volt}$$ $$C_{NV}^{yy}= \SI[parse-numbers = false]{(2.36\pm 0.12)\times 10^{-16}}{\newton / \volt}$$ An example of one of the PSDs used for calibration is presented in Fig. \ref{fig:calibration_panel}b).

\subsection{Gain matrix}

After ensuring proper calibration of the detectors and actuators, computation of the LQR gains becomes feasible. Analysis of the PSDs of the $x$, $y$ and $z$ confirms the trapped nanoparticle's oscillation frequencies $\Omega_x/2\pi=\SI[parse-numbers=false]{96.24}{k\hertz}$, $\Omega_y/2\pi=\SI[parse-numbers=false]{101.49}{k\hertz}$ and $\Omega_z/2\pi=\SI[parse-numbers=false]{31.52}{k\hertz}$. Given the average diameter of the nanoparticle as provided by the manufacturer, the its mass is calculated to be $m\approx\SI[parse-numbers]{3.37}{f\gram}$. The weighting and cost-effort matrices used were
\begin{equation}
\mathbf{R}_d = m\begin{bmatrix}
    \textrm{diag}(\mathbf{\Omega}^2) & \mathbf{0}_{3\times3}\\
    \mathbf{0}_{3\times3} & I_3\\
\end{bmatrix},
\end{equation}
and
\begin{equation}
    \mathbf{Q}_d=\frac{100}{m}\begin{bmatrix}
    \Omega_x^{-2}& 0 & 0 \\
    0 & \Omega_y^{-2} & 0 \\
    0 & 0 & \Omega_z^{-2} \\
\end{bmatrix}.
\end{equation}

These matrices were selected to ensure that the cost function $J_d$ possesses appropriate energy units, considering the states measured in S.I units and $\mathbf{u}$ accounting for feedback forces. Such dimensional considerations are crucial for converting the controller's gain from the LQR theory to the digital gains configured in the FPGA. The $\mathbf{B}$ matrix is expressed as 
\begin{equation}
\mathbf{B} = \begin{bmatrix}
    \mathbf{0}_{3\times 3}\\
    \mathbf{B}_{xyz}\\
\end{bmatrix}\textrm{, where } \mathbf{B}_{xyz}=\begin{bmatrix}
    \mathbf{B}_{xy} & 0\\
    \mathbf{0}_{1\times 2} & b_z
\end{bmatrix}.
\end{equation}

\begin{figure}[!t]
    \centering
    \includegraphics{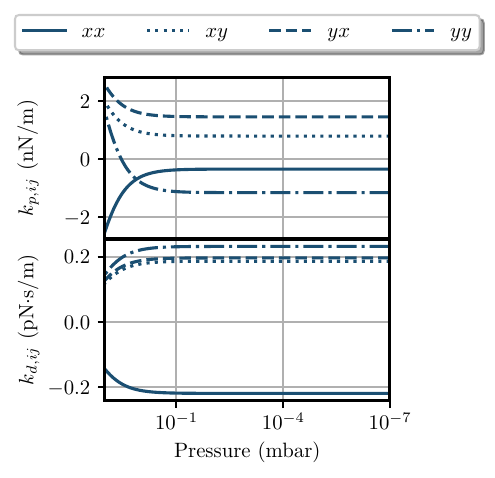}
    \caption{Optimal gains dependence with pressure. The constant behavior for values bellow $\SI[parse-numbers=false]{1}{mbar}$ allows one to employ the same matrix $\mathbf{K}_d$ for the underdamped and undamped regimes.} 
    \label{fig:gains_vs_pressure}
\end{figure}

The submatrix $\mathbf{B}_{xy}$, expressed in $\SI[parse-numbers=false]{}{k\gram^{-1}}$, is determined by $m$ and the proportion of the electrodes coefficients $C_{NV}^{ij}$, 
\begin{equation}
\mathbf{B}_{xy} = \frac{1}{m}\begin{bmatrix}
    -1 & C_{NV}^{xy}/C_{NV}^{xx} \\
    C_{NV}^{yx}/C_{NV}^{xx} & C_{NV}^{yy}/C_{NV}^{xx}
\end{bmatrix}.
\end{equation}
Without loss of generality, its terms were normalized by the biggest transduction coefficient, $C_{NV}^{xx}$. The negative sign accounts for the orientation of the electrodes axes, $x^\prime$ and $y^\prime$, as illustrated in Fig. \ref{fig:setup_simplified}.

The final parameter required to fully describe the dynamics given by Eq. \eqref{eq:2.state-variable-linear-system}, is $\gamma_m$. To assess the impact of varying it, we substitute the values for the resonance frequencies, $\mathbf{B}$, $T_s$, $\mathbf{Q}_d$, and $\mathbf{R}_d$ and compute $\mathbf{K}_d$ for different drag coefficients. The results of this evaluation are depicted in Fig. \ref{fig:gains_vs_pressure}. Notably, for pressures below $\SI[parse-numbers=false]{1}{mbar}$, the influence of $\gamma_m$ on the controller's gains is negligible. Therefore, under the premise that pressure solely affects the drag coefficient, $\mathbf{K}_d$ can be computed only once, even as pressure reduces.

After completing the system characterization, with $\gamma_m$ considered as zero, $\mathbf{K}_d$ can be properly computed. The next step involves converting the theoretical gains into digital values configured within the FPGA. The following expressions govern this conversion
\begin{subequations}\label{eq:digital-gains-conversion}
    \begin{align}
        k_{p,ij}^{d} &= \frac{k_{p,ij}}{AC_{NV}^{xx}C_{Vm}^j},\\
        k_{d,ij}^{d} &= -\frac{\Omega_j k_{d,ij}}{AC_{NV}^{xx}C_{Vm}^j}.
    \end{align}    
\end{subequations}
Here, $\Omega_j$ emerges from estimating the velocity as proportional to the delayed position, leading, for example, to $\dot{x}=-\Omega_x x(t-\tau_x)$, for a delay $\tau_x$. The factor $C_{NV}^{xx}$ arises from the $\mathbf{B}_{xy}$ matrix normalization, while $C_{Vm}^j$ is used to convert displacement in the $j$-axis to output voltages from its detector. In Table \ref{gain_values}, both theoretical and digital gains are presented. 

\begin{table}[!htb]
\begin{tabular}{ccc}
\hline \hline
Gain   &  LQR (Eq. \ref{eq:discrete-control-law}) & Digital Gains (Eq. \ref{eq:digital-gains-conversion})  \\ \hline
$k_{p,xx}$  & $\SI[parse-numbers=false]{-3.40 \times 10^{-10}}{N/m}$ & $-0.35$  \\
$ k_{p,xy}$  & $\SI[parse-numbers=false]{\phantom{+}7.99 \times 10^{-10}}{N/m}$ & $\phantom{+}0.80$  \\
$ k_{p,yx}$  & $\SI[parse-numbers=false]{\phantom{+}1.46 \times 10^{-9\phantom{0}}}{N/m}$ & $\phantom{+}1.50$  \\
$ k_{p,yy}$  & $\SI[parse-numbers=false]{-1.15 \times 10^{-9\phantom{0}}}{N/m}$ & $-1.15$  \\
$k_{d,xx}$  & $\SI[parse-numbers=false]{-2.19 \times 10^{-13}}{\newton\cdot\second/\meter}$ & $\phantom{+}136.45$  \\
$ k_{d,xy}$  & $\SI[parse-numbers=false]{\phantom{+}1.86 \times 10^{-13}}{\newton\cdot\second/\meter}$ & $-119.14$  \\
$ k_{d,yx}$  & $\SI[parse-numbers=false]{\phantom{+}1.96 \times 10^{-13}}{\newton\cdot \second/\meter}$ & $-122.22$  \\
$ k_{d,yy}$  & $\SI[parse-numbers=false]{\phantom{+}2.32 \times 10^{-13}}{\newton\cdot\second/\meter}$ & $-148.23$  \\
\hline \hline 
\end{tabular}
\caption{Controller's gains. Values returned by optimal control theory and implemented values within the FPGA are shown according to the system characterization and Eq. \eqref{eq:digital-gains-conversion}. The digital gains had to pass through a conversion to a fixed-point representation during the VHDL implementation,
allowing arithmetical operations with minimal loss of  numerical resolution \cite{wilson2015design}.}
\label{gain_values}
\end{table}

\end{document}